\documentclass[10pt,twocolumn,aps,pra]{revtex4-1} 

\usepackage{amsthm}
\usepackage{gensymb}
\usepackage{amsmath}            
\usepackage{amssymb}            
\usepackage{mathtools}
\usepackage{graphicx}           
\usepackage{xcolor}
\usepackage{braket}
\usepackage{etoolbox}
\usepackage{breqn}
\usepackage{epstopdf}
\usepackage{multirow}

\usepackage{xcolor}
\usepackage{bm}
\usepackage[normalem]{ulem}

\makeatletter
\let\cat@comma@active\@empty

\usepackage[colorlinks,citecolor=blue,urlcolor=blue,bookmarks=false,hypertexnames=true]{hyperref} 

\newcommand{\proj}[1]{\ket{#1}\bra{#1}}
\newcommand{\ts}{\otimes}
\newcommand{\id}{\openone}

\newcommand{\comment}[1]{}

\theoremstyle{plain}

\begin{document}
\title{Quick charging of a quantum battery with superposed trajecotries}

\author{Po-Rong Lai}
\affiliation{Department of Physics and Center for Quantum Frontiers of Research \&
Technology (QFort), National Cheng Kung University, Tainan 701, Taiwan}

\author{Jhen-Dong Lin}
\affiliation{Department of Physics and Center for Quantum Frontiers of Research \&
Technology (QFort), National Cheng Kung University, Tainan 701, Taiwan}

\author{Yi-Te Huang}
\affiliation{Department of Physics and Center for Quantum Frontiers of Research \&
Technology (QFort), National Cheng Kung University, Tainan 701, Taiwan}

\author{Yueh-Nan Chen}
\email{yuehnan@mail.ncku.edu.tw}
\affiliation{Department of Physics and Center for Quantum Frontiers of Research \&
Technology (QFort), National Cheng Kung University, Tainan 701, Taiwan}

\begin{abstract}
    We propose novel charging protocols for quantum batteries based on quantum superpositions of trajectories. Specifically, we consider that a qubit (the battery) interacts with multiple cavities or a single cavity at various positions, where the cavities act as chargers. Further, we introduce a quantum control prepared in a quantum superposition state, allowing the battery to be simultaneously charged by multiple cavities or a single cavity with different entry positions. To assess the battery's performance, we evaluate the maximum extractable work, referred to as ergotropy. Our main result is that the proposed protocols can utilize quantum interference effects to speed up the charging process. For the protocol involving multiple cavities, we observe a substantial increase in ergotropy as the number of superposed trajectories increases. In the case of the single-cavity protocol, we show that two superposed trajectories (entry positions) are sufficient to achieve the upper limit of the ergotropy throughout the entire charging process. Furthermore, we propose circuit models for these charging protocols and conduct proof-of-principle demonstrations on IBMQ and IonQ quantum processors. The results validate our theoretical predictions, demonstrating a clear enhancement in ergotropy.
\end{abstract}

\maketitle

\section{Introduction}
Quantum batteries (QBs) have emerged as a popular research topic, providing valuable insights into how thermodynamics functions at the quantum scale~\cite{Uzdin15,Niedenzu18}. Recent development has demonstrated that various quantum resources such as entanglement~\cite{Horodecki09,Bruß02} or coherence~\cite{Streltsov17,Nairz03} can enhance the performance of quantum batteries in terms of charging~\cite{Binder2015,Campaioli2017,Ferraro2018,Le18,Rossini19,Crescente20,Crescente20-2,Ghosh21}, storage~\cite{Santos2019,Primoradian19,Liu19,Quach2020,Gherardini2020,Rosa2020} and work extraction~\cite{Alicki2013,Francica2017,Manzano18,Andolina2019,Barra19,Francica20,Kamin20-2,Monsel20,Ghosh20}, etc. One of the intriguing phenomena used to achieve these enhancements is the collective effects triggered by a group of QBs~\cite{Mei92,Higgins2014,Campaioli2017,Ferraro2018,Primoradian19,Quach22,Joshi22}. In his seminal paper~\cite{Dicke54}, Dicke characterized one of the collective effects, superradiance, by the quantum interference of emissions from an ensemble of atoms. Recent investigations have also demonstrated the utility of the time-reversed phenomenon, known as superabsorption~\cite{Higgins2014,Yang21,Ueki2022,Quach22}, on enhancing the capabilities of QBs.

In this work, our focus lies on an interferometric approach known as ‘‘superpositions of trajectories"~\cite{Chiribella19,Foo20,Kristjansson20,Duprey22,Ghafari19,Rubino21,Foo21,Chan22,Lin2022,lin2023boosting,Ku2023coheret,Lee2023steering}. This approach treats an atom's space-time trajectories  as a quantum system, enabling the exploration of quantum interference of these trajectories. A notable outcome of this approach is the effective noise mitigation in various quantum information tasks~\cite{Ghafari19,Rubino21,Chan22,Ku2023coheret,Lee2023steering}. In our recent work~\cite{Lin2022}, we have further advanced the understanding by interpreting this noise mitigation as a Zeno-like state freezing phenomenon~\cite{misra1977zeno} within the framework of open quantum systems~\cite{breuer2002theory}. Additionally, we have demonstrated that this approach can also manifest Dicke-like collective effects even when only one single atom is involved. Building upon these insights, the present work aims to delve into the potential of leveraging superpositions of trajectories to enhance the performance of QBs.

We consider a qubit acting as the quantum battery, gaining energy through interactions with cavities functioning as chargers. To assess the QB's performance, we focus on the maximum extractable work, known as ergotropy~\cite{Alicki2013}. The ergotropy is bounded by the stored energy, i.e., the change in the qubit's internal energy, which can be regarded as a consequence of energy conservation. To utilize superposed trajectories, we propose two charging protocols using an interferometric setup akin to scenarios in Ref.~\cite{Lin2022}. The first one is called the multiple-charger protocol, which consists of multiple identical chargers (cavities), and the QB can interact with these chargers in a manner of quantum superposition via a multi-port beam splitter. In principle, there are multiple output beams of QB when it exits the interferometer. This enable us to adjust the work extraction strategy for each output and obtain an average ergotropy, also known as the daemonic ergotropy~\cite{Francica2017}. The primary result of this protocol is an ‘‘activation" of the ergotropy. Specifically, we demonstrate that when the battery is charged by a single cavity (without utilizing superposed trajectories), the ergotropy remains zero for a certain period, despite storing energy immediately after interaction with the cavity. Thus, there exists a finite delay before the battery can store ‘‘useful energy", i.e., extractable work. According to the definition of ergotropy, population inversion, i.e., the excited state population of the qubit being larger than its ground state population, is required to obtain a nontrivial ergotropy. Therefore, one must wait until the battery reaches the inversion point in order to obtain extractable work. 

Remarkably, we demonstrate that by considering this multiple-charger protocol, non-zero ergotropy can be obtained right after the charging process begins. This implies that this protocol enables the achievement of “quick charging" for the QB, where the ergotropy can be activated before reaching the inversion point. Furthermore, we observe that the ergotropy increases as the number of superposed trajectories grows. In the limit of an infinite number of superposed trajectories and considering the rotating-wave approximation, the ergotropy even saturates to its upper bound (i.e., the stored energy) throughout the charging process. This saturation indicates a complete conversion of stored energy into extractable work. In addition, we reveal that the enhancement of the average ergotropy stems from the increased average purity of the QB's output states, which can be regarded as a manifestation of the Zeno-like phenomenon described in Ref.~\cite{Lin2022}.

The second protocol is coined the single-charger protocol with only one cavity (charger) involved. In this protocol, the QB can enter from different positions into the cavity, experiencing different coupling strengths with the cavity. By using the superposed trajectories, the QB can enter these various positions simultaneously. As indicated in Ref.~\cite{Lin2022}, this particular setup can induce the collective interference effect. It is important to note that both the magnitude and relative phases of the coupling strengths between the QB and the charger can influence the collective interference effect and, consequently, the ergotropy. We show that two superposed trajectories (positions) are sufficient to reach the upper limit of the ergotropy, i.e., the stored energy, throughout the entire charging process. This is achievable if the two coupling strengths possess the same magnitude and are completely out of phase. Our analytical analysis demonstrates that this phenomenon originates from the collective constructive and destructive quantum interferences, which ensure that the output states remain pure throughout the charging process.

Additionally, we present quantum circuits designed for the aforementioned charging protocols, requiring fewer than twenty two-qubit gates. We implement and execute these circuits on both IonQ quantum processors (based on trapped ions) and IBMQ quantum processors (based on superconducting circuits). The experimental results obtained from these implementations further validate the increase in ergotropy, which is consistent with our theoretical predictions. 

The rest of the paper is organized as follows. In Sec.~\ref{independent}, we characterize the multiple-charger protocol. In Sec.~\ref{same}, we further investigate the single-charger protocol. In Sec.~\ref{device}, we consider the circuit implementations and present the experimental results of the devices from IBMQ and IonQ . Finally, we draw our conclusions in Sec.~\ref{summary}.

\section{Multiple-charger protocol}\label{independent}

\begin{figure}[!htbp]
\includegraphics[width = 1\columnwidth]{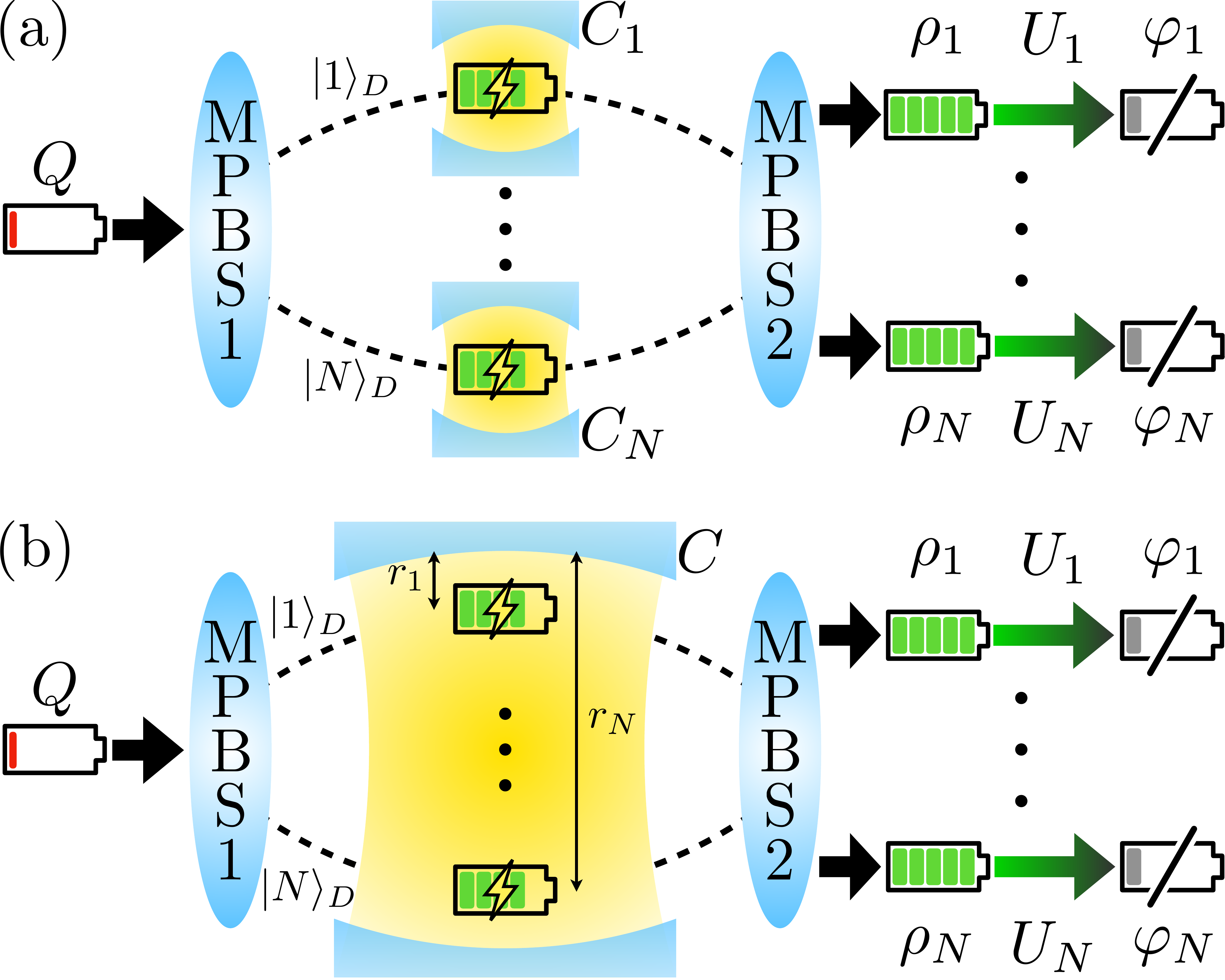}
\caption{The Quantum battery $Q$ is first sent into a multi-port beam splitter (MPBS1), which allows the quantum battery to travel along $N$ different trajectories (denoted by $\ket{j}_D$, where $j=1\cdots N$ in the following) in a manner of quantum superposition. We consider two charging processes: (a) the trajectories each lead to a charger (cavities) $\{C_j\}$, causing the quantum battery to interact with all chargers simultaneously, (b) the trajectories lead to a single charger $C$ but at different positions $\{r_j\}$, causing the quantum battery to interact with the charger with various coupling strengths. Once the charging process is completed, a second multi-port beam splitter (MPBS2) is used to perform measurement on the trajectories degree of freedom $D$. This measurement captures the quantum interference effect between different trajectories and results in $N$ possible reduced states $\rho_{j}$. We then extract work from each $\rho_j$, where the maximum amount of extractable work is called the ergotropy. The work extraction operations are described by unitary operators $U_j$, which transforms each of the batteries to a passive state $\varphi_{j}$.}
\label{setup process}
\end{figure}

We now formulate the multiple-charger scenario, which can be described by an interferometric setup as shown in Fig.~\ref{setup process}(a). The charging protocol consists of three different components: (i) a qubit $Q$, which acts as the quantum battery. (ii) $N$ identical single-mode cavities $\left\{C_j\right\}_{j=1\cdots N}$, which act as the chargers. The QB moves at a speed $v$ and gets charged when it passes through one of the chargers. Suppose that the cavity length is $l$. Then, the interaction time reads $\tau = l/v$. To simplify our discussions, we assume that the cavity is homogeneous such that the interaction strength between the QB and the charger remains constant during the charging process~\cite{Meschede85}. (iii) We characterize the trajectory degrees of freedom as an $N$ dimensional qudit $D$, wherein we associate $N$ different trajectories inside the interferometer with $N$ basis states $\left\{\ket{j}_D\right\}_{j=1\cdots N}$. When the QB takes the path labeled by $j$, it interacts with the charger $C_j$. In other words, $D$ acts as a quantum control that determines which charger the QB interacts with. The total Hamiltonian involving these three components can then be written as
\begin{equation}\label{hamiltonian}
    H_{\text{tot}}=\sum_{j=1}^N \ket{j}\bra{j}_D\otimes H_{Q C_j}.
\end{equation}
The Hamiltonian $H_{Q C_j}$ of the quantum battery $Q$ and the charger $C_j$ can be expressed as
\begin{equation}\label{eq:H_QCj}
\begin{aligned}
    H_{Q C_j}&=H_Q+H_{C_j}+H_{QC_j}',\\
    H_Q&= 
    \frac{\hbar}{2}\omega_a \hat{\sigma}_{z}=\frac{\hbar}{2}\omega_c \hat{\sigma}_{z},\\
    H_{C_j}&=\hbar\omega_c \hat{a}_{j}^{\dag} \hat{a}_{j},\\
    H_{QC_j}'&=\hbar\omega_c\lambda\hat{\sigma}_x \left(\hat{a}_{j} + \hat{a}_{j}^{\dag} \right).
\end{aligned}
\end{equation}
Here, $\hat{a}_{j}\left(\hat{a}_{j}^{\dag}\right)$ annihilates (creates) a photon in $C_j$ with frequency $\omega_c$, $\hbar\omega_a$ represents the energy splitting between the ground state $\ket{g}$ and the excited state $\ket{e}$ of $Q$. The Pauli operators are therefore given by $\hat{\sigma}_z=\ket{e}\bra{e}-\ket{g}\bra{g}$ and $\hat{\sigma}_x=\ket{e}\bra{g}+\ket{g}\bra{e}$. Moreover, the dimensionless constant $\lambda$ represents the coupling strength between $Q$ and all the chargers. Throughout this work, we focus on the resonant regime $\omega_a = \omega_c$.

We first send the battery $Q$ into the multiport beam splitter (MPBS1 in Fig.~\ref{setup process}). In general, the beam splitter can prepare the trajectories in a quantum superposition state so that $Q$ can be charged by these $N$ chargers simultaneously. For simplicity, we assume the superposition state of the trajectories is
\begin{equation}
    \ket{\psi}_D=\frac{1}{\sqrt{N}}\sum_{i=1}^N \ket{j}_D.
\end{equation}
We consider that the battery and chargers are initialized in the ground state and the single-photon Fock state, respectively. Therefore, the total initial state reads
\begin{equation}\label{initial_independent}
    \ket{\psi\left(0\right)}_{DQC}=\frac{1}{\sqrt{N}}\sum_{j=1}^N\ket{j}_D\ts\ket{g}_Q\bigotimes_{j=1}^N\ket{1}_{C_j}.
\end{equation}
After $Q$ interacts with the chargers, according to Eq.~\eqref{hamiltonian}, the total states becomes
\begin{equation}
    \ket{\psi\left(\tau\right)}_{DQC}=\frac{1}{\sqrt{N}}\sum_{j=1}^N\ket{j}_D\ts\ket{\phi_{j}\left(\tau\right)}_{QC},
\end{equation}
where $\ket{\phi_{j}\left(\tau\right)}_{QC}$ is defined as
\begin{equation}\label{eq:Phi_j}
    \ket{\phi_{j}\left(\tau\right)}_{QC}=\exp\left(-i\frac{\tau}{\hbar} H_{QC_j}\right)\left(\ket{g}_Q\bigotimes_{j=1}^N\ket{1}_{C_j}\right).
\end{equation}

Finally, we make these trajectories interfere with one another by using another beam splitter (MPBS2 in Fig.~\ref{setup process}). In principle, MPBS2 has $N$ different outputs, which can be described using a set of orthonormal projectors $\{P_k\}_k$ acting on $D$, namely
\begin{equation}\label{eq:projector}
\begin{aligned}
    &P_{k}=\ket{\xi_k}\bra{\xi_k}_{D},\\ 
    &\sum_{k = 1}^{N} P_{k}=\openone,\\
    &\braket{\xi_k|\xi_{k'}} =\delta_{k,k'}~\forall~k,k'.
\end{aligned}
\end{equation}
Therefore, the (unnormalized) reduced state for the system $Q$ with the output $k$ reads 
\begin{equation}\label{independent final eq}
\begin{aligned}
\sigma_{k}\left(\tau\right)= \text{Tr}_{CD}\left[P_{k}\ket{\psi\left(\tau\right)}\bra{\psi\left(\tau\right)}_{DQC}P_{k}\right].
\end{aligned}
\end{equation}
Note that the probability of obtaining the outcome $k$ is  $p_{k}(\tau)=\mathrm{Tr}\left[\sigma_{k}(\tau)\right]$. Thus, the normalized state conditioned on the outcome $k$ can be written as $\rho_{k}(\tau)= \sigma_{k}(\tau)/p_{k}$.

Throughout this work, we choose
\begin{equation}\label{eq:projector2}
\begin{aligned}
    &\ket{\xi_{k=1}}\bra{\xi_{k=1}}_{D} \equiv \frac{1}{N}\sum_{m,n=1}^N\ket{m}\bra{n}_D.
\end{aligned}
\end{equation}
According to the assumption that all chargers are identical as well as the orthonormality of the projectors, in Appendix~\ref{irrelevant projectors}, we show that the explicit form of the rest of the projectors is irrelevant, enabling us to further simplify the analysis.

Here, we evaluate the performance of a QB by considering the ergotropy, which quantifies the maximum extractable work. Given a charged state of the QB $\rho(\tau)$, the ergotropy is defined as
\begin{equation}
\begin{aligned}
    W(\rho(\tau)) &\equiv \text{Tr}\left(\rho(\tau) H_Q\right) - \min_U~\text{Tr}\left(U\rho(\tau) U^{\dag} H_Q\right) \\
    &= \text{Tr}\left(\rho(\tau) H_Q\right) - \text{Tr}\left(\varphi(\tau) H_Q\right), \label{eq:ergotropy}
\end{aligned}
\end{equation}
where $U$ represents the unitary operation for work extraction. Also, $\varphi$ is known as the passive state~\cite{Alicki2013} (associated with $\rho$), which cannot provide useful work for all possible work extraction operations $U$. According to Ref.~\cite{Alicki2013}, the passive state of the battery can be written as 
\begin{equation}
    \begin{aligned}
        \varphi = s_0 \ket{e}\bra{e} + s_1\ket{g}\bra{g}.
    \end{aligned}
\end{equation}
Here, $s_0$ and $s_1$ denote the eigenvalues of $\rho$ with $s_0<s_1$.
Note that the upper limit of the ergotropy is set by the stored energy quantified by the difference in the internal energy of the QB before and after charging, namely 
\begin{equation}
\begin{aligned}
    E\left(\tau\right) &\equiv\sum_{k=1}^N p_{k}\left(\tau\right)\text{Tr}\left[H_Q \rho_{k}\left(\tau\right)\right]- \text{Tr}\left(H_Q\ket{g}\bra{g}\right)\\
    &=\text{Tr}\left[H_Q \rho(\tau)\right] - \text{Tr}\left(H_Q\ket{g}\bra{g}\right),
\end{aligned}
\end{equation}
where $\rho(\tau) = \sum_{k}\sigma_{k}(\tau)$.
We can obtain $W\leq E$, because $\mathrm{Tr}(H_Q\ket{g}\bra{g}) \leq \mathrm{Tr}(H_Q \varphi)$ in general. In addition, the inequality saturates if and only if $\varphi  = \proj{g}$, which implies that $\rho$ is a pure state.

As aforementioned, in our charging protocol, there are $N$ different outputs (labeled as $\{k\}$). In principle, one can find the optimal work extraction strategies for each output and obtain the average ergotropy, i.e.,
\begin{equation}
    \begin{aligned}
        \overline{W} &= \sum_{k} p_{k}W(\rho_{k}(\tau)) \\
        &= \sum_{k}p_{k}\left[\mathrm{Tr}\left(\rho_{k}(\tau) H_Q\right)-\mathrm{Tr}\left(\varphi_{k}(\tau) H_Q\right)\right],
    \end{aligned}
\end{equation}
where $\varphi_{k}$ denotes the passive state associated with $\rho_{k}$. Following similar reasoning as mentioned earlier, the average ergotropy is also upper bonded by the stored energy, and the optimal extractable work can be obtained, i.e., $\overline{W}(\tau) = E(\tau)$, if and only if $\{p_k,\rho_{k}\}$ forms a pure state decomposition of $\rho$.

To gain some analytical insight, we now consider the rotating-wave approximation, which is usually valid for coupling strength $\lambda\leq 0.1$~\cite{FriskKockum2019}, so that the interaction Hamiltonian of the QB and the chargers in Eq.~(\ref{eq:H_QCj}) can be reduced to the Jaynes-Cummings model, namely
\begin{equation}
    \tilde{H}_{Q C_j}'=\hbar\omega_c\lambda \left( \hat{\sigma}_{+}\hat{a}_{j} + \hat{\sigma}_{-}\hat{a}_{j}^{\dag} \right),
\end{equation}
where $\hat{\sigma}_{+}=\ket{e}\bra{g}$ and $\hat{\sigma}_{-}=\ket{g}\bra{e}$ represent the creation   and annihilation operators of $Q$, respectively. We can then evaluate Eq.~(\ref{eq:Phi_j}) in this case:
\begin{equation}
\begin{aligned}
    \ket{\phi_{j}\left(\tau\right)}_{QC}=&-i\sin\left(\omega_c\lambda\tau\right)\ket{e}_Q\ts\hat{a}_{j}\bigotimes_{j'=1}^N\ket{1}_{C_{j'}} \\ 
    &+ \cos\left(\omega_c\lambda\tau\right)\ket{g}_Q\bigotimes_{j=1}^N\ket{1}_{C_j}.
\end{aligned}
\end{equation}
Let us start from the simplest case with only one charger (i.e., $N=1$), where the reduced state of $Q$ is expressed as \begin{equation}
    \rho(\tau) = \sin^2\left(\omega_c\lambda\tau\right)\ket{e}\bra{e}+\cos^2\left(\omega_c\lambda\tau\right)\ket{g}\bra{g}.
\end{equation} In this case, the stored energy is 
\begin{equation}
E(\tau)=\hbar\omega_c\sin^2\left(\omega_c\lambda\tau\right),
\end{equation}
which oscillates with a period $T = 2\pi/(\omega_c \lambda) $. We now focus on the time interval $\tau \in [0,T/4]$ (such that $\omega_c\lambda\tau\in[0,\pi/2]$), where the stored energy monotonically increases from $0$ to its maximum value $\hbar \omega_c$. Note that $\rho$ is diagonalized under basis $\{\ket{e},\ket{g}\}$. Thus, according to Eq.~\eqref{eq:ergotropy}, the criterion for obtaining non-zero ergotropy is the moment that population inversion occurs, where the excited state population becomes larger than the ground state population ($\bra{e}\rho(\tau)\ket{e} > \bra{g}\rho(\tau)\ket{g}$). The time-dependence of the ergotropy can then be derived as 
\begin{equation}\label{eq:N=1case}
    W(\tau)= 
    \begin{cases}
        0 & \textrm{if}~0\leq\tau<\frac{T}{8},\\
        \hbar\omega_c (2\sin^2\left(\omega_c\lambda\tau\right)-1) & \textrm{if}~\frac{T}{8}\leq\tau\leq\frac{T}{4}.
    \end{cases}
\end{equation}
One can observe that in the duration $\tau \in [0,T/8]$, although the stored energy $E$ monotonically increases, there is no extractable work, $W = 0$, for the battery because $\rho$ remains a passive state during this period.

\begin{figure}[!htbp]
\includegraphics[width=1\columnwidth]{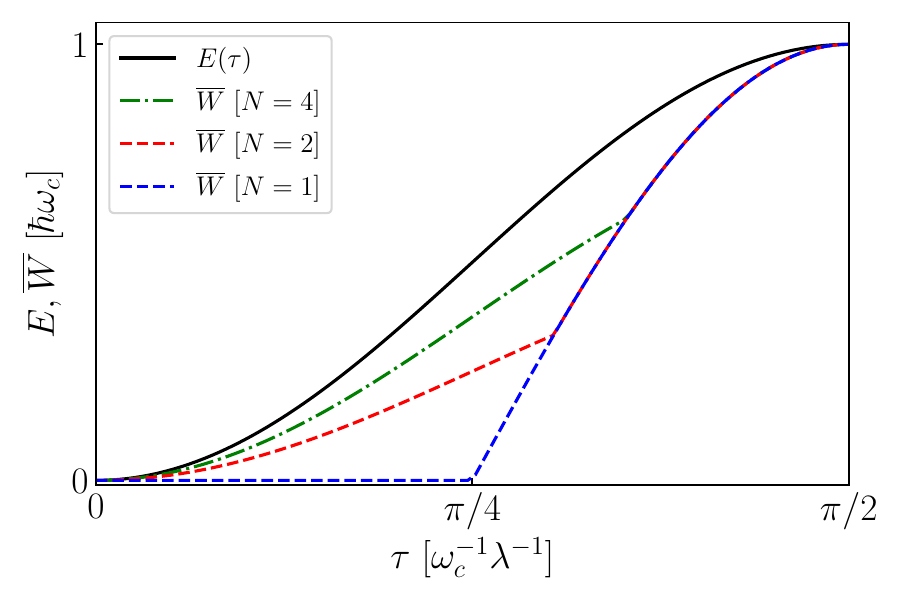}
\caption{The stored energy $E$ and average ergotropy $\overline{W}$ (both in units of $\hbar \omega_c$) on time $\tau$ (in units of $1/\omega_c$) for $\lambda = 0.05$. The black solid curve plots the stored energy while the dashed curves plot the average ergotropy. From bottom to top, the blue, red and green dashed curves show the results for $N=1$, $N=2$ and $N=4$, respectively.}
\label{jcindependent result}
\end{figure}

We now consider the scenario involving $N$ chargers. When the selective measurements satisfy Eq.~(\ref{eq:projector}) and Eq.~(\ref{eq:projector2}), the unnormalized post-measurement states can be written as (see Appendix~\ref{irrelevant projectors} for detailed derivations):
\begin{equation}\label{eq:id post states}
\begin{aligned}
    \sigma_{k=1}\left(\tau\right)=&\frac{1}{N}\sin^2\left(\omega_c\lambda\tau\right)\ket{e}\bra{e} + \cos^2\left(\omega_c\lambda\tau\right)\ket{g}\bra{g}, \\
    \sigma_{k\neq1}\left(\tau\right)=&\frac{1}{N}\sin^2\left(\omega_c\lambda\tau\right)\ket{e}\bra{e}.
\end{aligned}
\end{equation}
Here, for the case $k=1$, the post-measurement state is passive during the time period $\tau\in \left[0, T_N\right]$ with the inversion time $T_N = \tan^{-1}(\sqrt{N})T/2\pi$. Remarkably, for the cases of $k \neq 1$, the post-measurement states are exactly the excited state, implying that the maximal extractable work $\hbar \omega_c$ can be obtained. Therefore, the average ergotropy can be expressed as 
\begin{equation}\label{eq:jc energy}
\begin{aligned}
    &\overline{W}\left(\tau\right)\\
    &=W(\sigma_{k=1}(\tau))+(N-1)W(\sigma_{k\neq1}(\tau))\\
    &=\begin{cases}
        \hbar\omega_c\frac{N-1}{N}\sin^2\left(\omega_c\lambda\tau\right) & \text{if}~0\leq \tau \leq T_N\\
        \hbar\omega_c\left(2\sin^2\left(\omega_c\lambda\tau\right)-1\right) & \text{if}~T_N\leq \tau \leq \frac{T}{4}.
    \end{cases}    
    \end{aligned}
\end{equation}
In Fig.~\ref{jcindependent result}, we present the time-dependent stored energy and the average ergotropy for different values of $N$. In contrast to the case of $N=1$, we observe non-zero average ergotropy for the entire interval of interest because the states with $k \neq 1$ are non-passive right after the QB-chargers interaction is turned on ($\tau>0$). Therefore, \textit{the protocol can be used for “quick charging", enabling immediate storage of useful work after the charging process begins}. Furthermore, the result indicates that increasing $N$ delays the inversion time $T_N$ and enhances the average ergotropy before $T_N$. According to Eq.~\eqref{eq:jc energy}, this enhancement originates from the increase of the purity for the output $k=1$ as its excited state population decreases with $N$. This result aligns with the Zeno-like state freezing effect described in Ref.~\cite{Lin2022}. In the asymptotic limit ($N\rightarrow \infty$), we can further obtain a pure state decomposition, i.e., $\sigma_{k=1}(\tau)\propto \proj{g}$ $\sigma_{k\neq1}(\tau)\propto \proj{e}$, implying that the stored energy can be fully converted into extractable work for the whole time interval, i.e, $E(\tau)=\overline{W}(\tau)$. 

\begin{figure*}[!htbp]
\includegraphics[width=2\columnwidth]{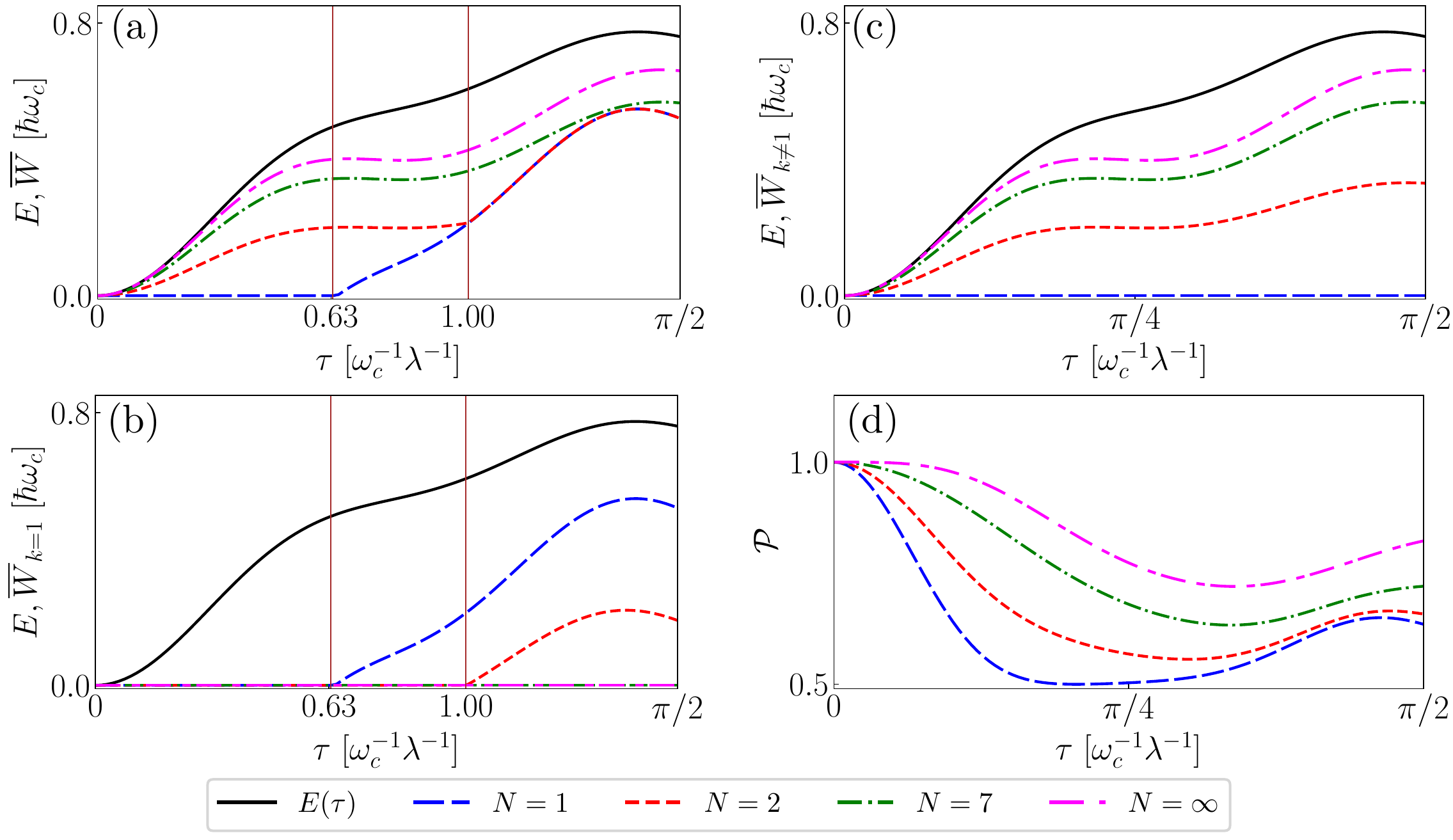}
\caption{(a)~The stored energy $E$ and average ergotropy $\overline{W}$ (both in units of $\hbar \omega_c$) as functions of time $\tau$ (in units of $1/\omega_c$) for $\lambda = 0.5$. (b)~The stored energy $E$ and the average ergotropy contributed by states of $k=1$ $\overline{W}_{k=1}$. Two brown vertical lines at $\tau=0.63,1.00$ indicate the inversion points $T_N$ for $N=1$(blue) and $N=2$(red). Here, $\overline{W}_{k=1}$ is 0 when $N=7$(green) and $N=\infty$(magenta). (c)~The stored energy $E$ and the average ergotropy contributed by states of $k\neq1$ $\overline{W}_{k\neq 1}$ on time $\tau$. Except for $N=1$ (blue), these states contribute average ergotropy when $\tau>0$. (d)~The dashed curves plot the change in average purity $\mathcal{P}$ against time $\tau$. The cutoff photon number is set to 9 in the above results.}
\label{ultraindependent result}
\end{figure*}

We now extend our scope of discussion into the ultrastrong--coupling regime~\cite{FriskKockum2019}, where the rotating--wave approximation is no longer valid. In Fig.~\ref{ultraindependent result}~(a), we present the dynamics of the average ergotropy and the stored energy. We can still observe the quick charging effect, a delay of the inversion point, and an enhancement in average ergotropy as $N$ increases. As shown in Figs.~\ref{ultraindependent result}~(b) and (c), we further present the individual contributions of the average ergotropy from the outputs $k=1$ and $k\neq 1$, which are respectively defined by 
\begin{equation}
    \begin{aligned}
        &\overline{W}_{k=1}=p_{k=1}W(\rho_{k=1}(\tau)) \\
        &\text{and} \\
        &\overline{W}_{k\neq1}=\sum_{k\neq1}p_k W(\rho_k(\tau)).
    \end{aligned}
\end{equation}
We can observe that the inversion points come from the contribution of $k=1$. Further, we can observe that its contribution decreases as $N$ increases. When $N\geq 7$, the contribution vanishes for the entire charging period. In Appendix~\ref{irrelevant projectors}, we provide analytical analysis, showing that the decrease in average ergotropy can also attribute to the Zeno-like state freezing effect. More specifically, we demonstrate that the excited state population decreases when $N$ increases. In the asymptotic limie ($N\rightarrow\infty$), the state can be frozen in the ground state. We further observe that the quick charging effect originates from the contribution of $k\neq 1$, since the corresponding average ergotropy becomes non-zero as soon as the charging process begins when $N> 2$. However, in contrast to the previous results with the rotating--wave approximation, the average ergotropy cannot reach the upper bound even in the asymptotic limit, implying that the post-measurement states are not pure. 

\begin{figure}[!htbp]
\includegraphics[width=1\columnwidth]{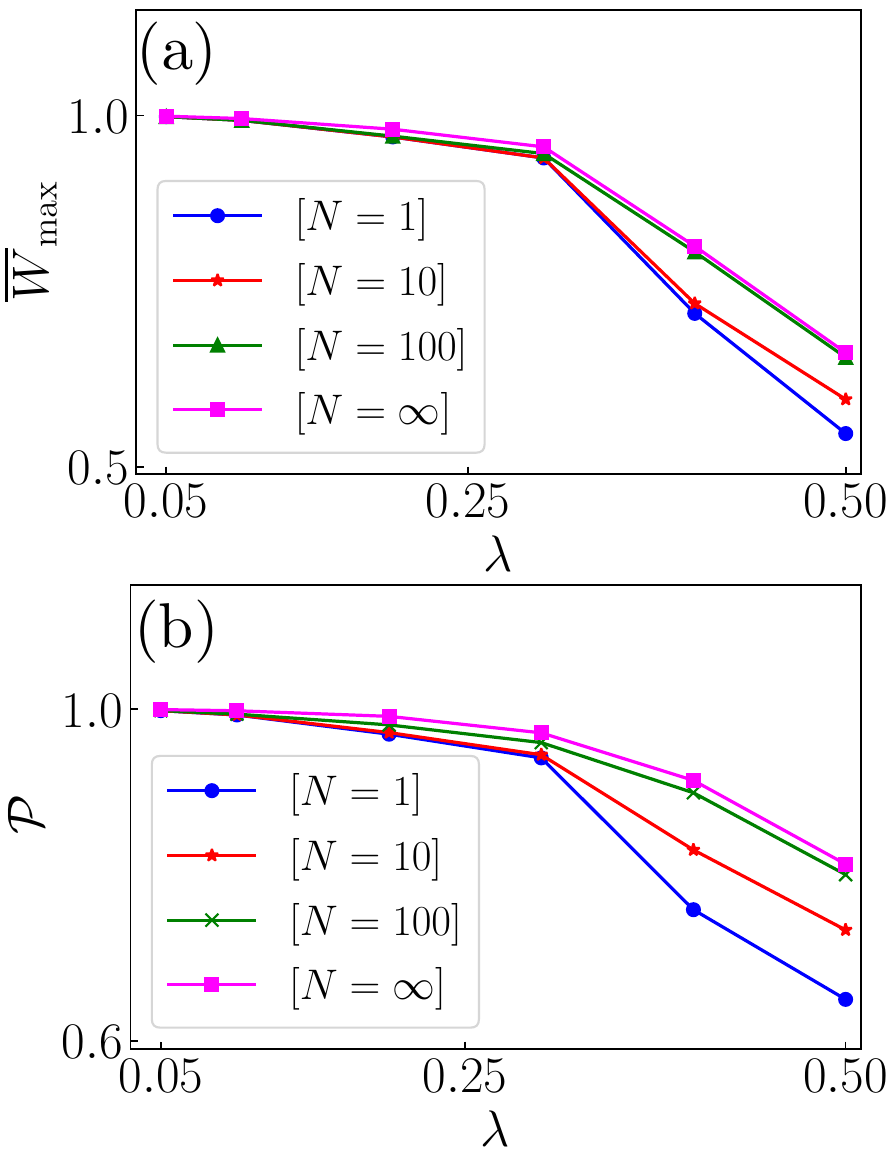}
\caption{(a)  Maximum average ergotropy $\overline{W}_{\textrm{max}}$ defined in Eq.~\eqref{eq:max ergotropy} with respect to different coupling strength $\lambda$. The blue, red, green and magenta dashed lines represent the results for $N=1$, $N=10$, $N=100$, and $N=\infty$, respectively. (b)The average purity $\mathcal{P}$ on dimensionless coupling strength $\lambda$. The blue, red, green and magenta dashed lines represent the results for $N=1$, $N=10$, $N=100$,
and $N=\infty$, respectively. The cutoff photon number is set to 9 in the above results. 
}
\label{fig:overlambda}
\end{figure}

In Fig.~\ref{ultraindependent result}(d), we present the average purity $\mathcal{P}$ associated with the post-measurement states, which is defined by
\begin{equation}
\begin{aligned}
    &\mathcal{P}=\sum_k \text{Tr}[\sigma_{k}]\text{Tr}\left[\left(\frac{\sigma_{k}}{\text{Tr}[\sigma_{k}]}\right)^2\right].
\end{aligned}
\end{equation}
We can observe the overall average purity increasing as $N$ increases, thereby leading to the enhancement of average ergotropy. However, the average purity cannot reach unity even in the asymptotic limit. Thus, the stored energy cannot be fully converted into extractable work. In Figs.~\ref{fig:overlambda} (a) and (b), we compare the maximum average ergotropy with the average purity with respect to the coupling strength $\lambda$, where the maximum average ergotropy is  defined by
\begin{equation}
    \overline{W}_{\textrm{max}}=\max_{\tau\in[0,T/4]}\overline{W}(\tau).\label{eq:max ergotropy}
\end{equation}
We can observe that as the coupling strength increases, the average purity drops, hence leading to a decrease in the maximum average ergotropy.

\section{Single-charger protocol}\label{same}

In this section, we discuss the single-charger protocol as shown in Fig.~\ref{setup process}(b), where the quantum battery can enter and interact with a single cavity (charger) $C$ at different positions.
The charging protocol resembles the one in Sec.~\ref{independent}, albeit with two minor adjustments. First, we only use a single charger $C$ instead of multiple ones. Second, the $N$ dimensional qudit $D$ now acts as a quantum control that decides the QB's position of entry into $C$. More specifically, a path labeled by $j$ guides the QB to a designated  position denoted as $r_j$ inside the cavity. 

In this scenario, the QB experiences a varying coupling strength with respect to the position inside that charger~\cite{Shanks2013}. Therefore, the position-dependent QB-charger Hamiltonian can be written as
\begin{equation}\label{eq:same hamiltonian}
    H_{j}=H_Q+H_C+H_j',
\end{equation}
where $H_C=\hbar\omega_c \hat{a}^\dagger \hat{a}$. Here, $H_j'$ represents the interaction Hamiltonian when QB is located at the position $r_j$. Its explicit form is given by
\begin{equation}
    H_j'=\hbar\omega_c\lambda\cos{\left(\pi\frac{r_j}{L}\right)}\hat{\sigma}_x \left(\hat{a} + \hat{a}^{\dag} \right),
\end{equation}
where $L$ denotes the width of the charger and $\lambda$ represents the maximal QB-charger coupling strength, which can be achieved when $r_j=0$ or $r_j=L$. Thus, the total Hamiltonian for this scenario can be expressed by
\begin{equation}\label{eq:total same hamiltonian}
    H_{\text{tot}}=\sum^N_{j=1}\ket{j}\bra{j}_D\ts H_j.
\end{equation}

Similar to the previous consideration, we prepare the total system in the following initial state: 
\begin{equation}
    \ket{\psi(0)}_{DQC}=\frac{1}{\sqrt{N}}\sum^N_{j=1}\ket{j}_D\ts\ket{g}_Q\ts\ket{1}_C,
\end{equation}
and allow it to evolve according to the total Hamiltonian $H_{\text{tot}}$, namely
\begin{equation}
    \ket{\psi(\tau)}_{DQC}=\frac{1}{\sqrt{N}}\sum^N_{j=1}\ket{j}_D\ts\ket{\phi_j(\tau)}_{QC},
\end{equation}
where $\ket{\phi_j(\tau)}_{QC}$, in this case, is defined as
\begin{equation}
    \ket{\phi_j(\tau)}_{QC}=\exp\left(-i\frac{\tau}{\hbar}H_j\right)\left(\ket{g}_Q\ts\ket{1}_C\right).
\end{equation}
We also consider the projectors defined in Eq.~(\ref{eq:projector}) to characterize the measurements performed by using MPBS2, such that the corresponding post-measurement states reads
\begin{equation}\label{eq:same post states}
    P_k\ket{\psi(\tau)}_{DQC}=\ket{\xi_k}_D\ts\sum^N_{j=1}c_{k,j}\ket{\phi_j(\tau)}_{QC},
\end{equation}
where the coefficient $c_{k,j}$ is given by
\begin{equation}\label{eq:c_k,j}
    c_{k,j}=\frac{1}{\sqrt{N}}\langle\xi_k|j\rangle.
\end{equation}

We now switch to the interaction picture, such that the interaction Hamiltonian associated with the position $r_j$ can be expressed as
\begin{equation}
\begin{aligned}
    H_j'(\tau)=&e^{\frac{i}{\hbar}(H_Q+H_C)\tau} ~H_j'~ e^{-\frac{i}{\hbar}(H_Q+H_C)\tau}\\
    =&\cos\left(\pi\frac{r_j}{L}\right)H_I'(\tau),
\end{aligned}
\end{equation}
where $H_I'(\tau)$ is the position-independent part and reads
\begin{equation}
    H_I'(\tau)=\hbar\omega_c\lambda e^{\frac{i}{\hbar}(H_Q+H_C)\tau} \hat{\sigma}_x(\hat{a}+\hat{a}^\dagger)e^{-\frac{i}{\hbar}(H_Q+H_C)\tau}.
\end{equation}
We can now characterize the time evolution with the propagator $U_{j,I}(\tau,0)$ in terms of the Dyson series, namely
\begin{equation}
\begin{aligned}
    U_{j,I}(\tau,0)=&\hat{\mathcal{T}}\exp\left[\frac{-i}{\hbar}\int_0^\tau H_j'(t')dt'\right] \\
    =&\sum_{n=0}^\infty\frac{1}{n!}\left[-\frac{i}{\hbar}\cos\left(\pi\frac{r_j}{L}\right)\right]^n\\
    &\times\int_0^\tau dt_1 \cdots \int_0^\tau dt_n \hat{\mathcal{T}} H_I'(t_1)\cdots H_I'(t_n),
\end{aligned}
\end{equation}
where $\hat{\mathcal{T}}$ is the time-ordering operator with $t_1>t_2>\cdots>t_n$. The post-measurement state of the QB and the charger in Eq.~(\ref{eq:same post states}) can then be expressed by
\begin{equation}\label{eq:evolution int. picture}
    \sum_{j=1}^N c_{k,j}\ket{\phi_j(\tau)}_{QC}=\sum_{j=1}^N c_{k,j} U_{j,I}(\tau,0)\ket{g}_{Q}\otimes\ket{1}_{C}.
\end{equation}
Therefore, one can observe that the total evolution (including the qudit, the QB, and the charger) is described by a linear combination of the position-dependent propagators, which characterizes the collective quantum interference effect among different positions~\cite{Lin2022}.

We now show that two superposed trajectories (positions) can lead to the saturation of the ergotropy to its upper bound with an appropriate adjustment of the collective interference effect. More specifically, we consider that the two positions satisfy $r_1+r_2=L$, such that
\begin{equation}\label{eq:cos_r with N=2}
    \cos\left(\pi\frac{r_2}{L}\right)=\cos\left(\pi-\pi\frac{r_1}{L}\right)=-\cos\left(\pi\frac{r_1}{L}\right).
\end{equation}
Therefore, the coupling strengths share the same magnitude but are completely out of phase.
Furthermore, we consider
\begin{equation}\label{eq:same projectors}
\begin{aligned}
    \ket{\xi_{k=1}}&=\frac{1}{\sqrt{2}}(\ket{1}_D+\ket{2}_D),\\
    \ket{\xi_{k=2}}&=\frac{1}{\sqrt{2}}(\ket{1}_D-\ket{2}_D),
\end{aligned}    
\end{equation}
and, according to Eq.~(\ref{eq:c_k,j}), the coefficients in this case are
\begin{equation}\label{eq:c_k,j with N=2}
    c_{1,1}=c_{1,2}=c_{2,1}=-c_{2,2}=\frac{1}{2}. 
\end{equation}
Therefore, the time evolution of the post-measurement state of the QB and the charger for $k=1$ is given by
\begin{equation}\label{eq:same k=1 post-measure}
\begin{aligned}
    \sum_{j=1}^2& c_{1,j}U_{j, I}(\tau,0)\ket{g}_{Q}\otimes\ket{1}_{C}\\
    =&\frac{1}{2}[U_{1,I}(\tau,0)+U_{2,I}(\tau,0)]\ket{g}_{Q}\otimes\ket{1}_{C}\\
    =&\sum_{n ~\textrm{even}}^\infty\frac{1}{n!}\left[-\frac{i}{\hbar}\cos\left(\pi\frac{r_1}{L}\right)\right]^n\\
    &\times\int_0^\tau dt_1 \cdots \int_0^\tau dt_n \hat{\mathcal{T}} H_I'(t_1)\cdots H_I'(t_n)\ket{g}_{Q}\otimes\ket{1}_{C}.
\end{aligned}
\end{equation}
We can observe that in the Dyson series, all the odd terms vanish, leading to the phenomenon of destructive interference. This effect originates from the complete out-of-phase nature of the coupling strengths for the two positions. As a direct consequence, the QB remains in the ground state $\ket{g}$ throughout the entire process.
Analogously, the post-measurement state for $k=2$ can be written as
\begin{equation}\label{eq:same k=2 post-measure}
\begin{aligned}
    \sum_{j=1}^2& c_{2,j}U_{j, I}(\tau,0)\ket{g}_{Q}\otimes\ket{1}_{C}\\
    =&\frac{1}{2}[U_{1,I}(\tau,0)-U_{2,I}(\tau,0)]\ket{g}_{Q}\otimes\ket{1}_{C}\\
    =&\sum_{n ~\textrm{odd}}^\infty\frac{1}{n!}\left[-\frac{i}{\hbar}\cos\left(\pi\frac{r_1}{L}\right)\right]^n\\
    &\times\int_0^\tau dt_1 \cdots \int_0^\tau dt_n \hat{\mathcal{T}} H_I'(t_1)\cdots H_I'(t_n)\ket{g}_{Q}\otimes\ket{1}_{C}.
\end{aligned}
\end{equation}
In this case, all the even terms in the Dyson series vanish, implying the QB is in the excited state $\ket{e}$ for all $\tau>0$ (with zero probability of obtaining the output $k=2$ at $\tau=0$).
Because the post-measurement QB states for these two outputs are pure states (i.e, the average purity is one), we can conclude that the stored energy can be fully converted to the extractable work, i.e., $E=\overline{W}$, throughout the whole charging process. Note that the presented analysis does not rely on the rotating wave approximation, thus indicating the saturation of the ergotropy to its upper limit generally holds for all regimes of the QB-charger coupling strength.

\section{Implementation on Quantum Devices}\label{device}
In this section, we provide circuit models for the proposed charging protocols and perform proof-of-concept experiments on the quantum processors provided by IBMQ and IonQ, which involves two superposed trajectories ($N=2$).

\begin{figure*}[!htbp]
\includegraphics[width=2\columnwidth]{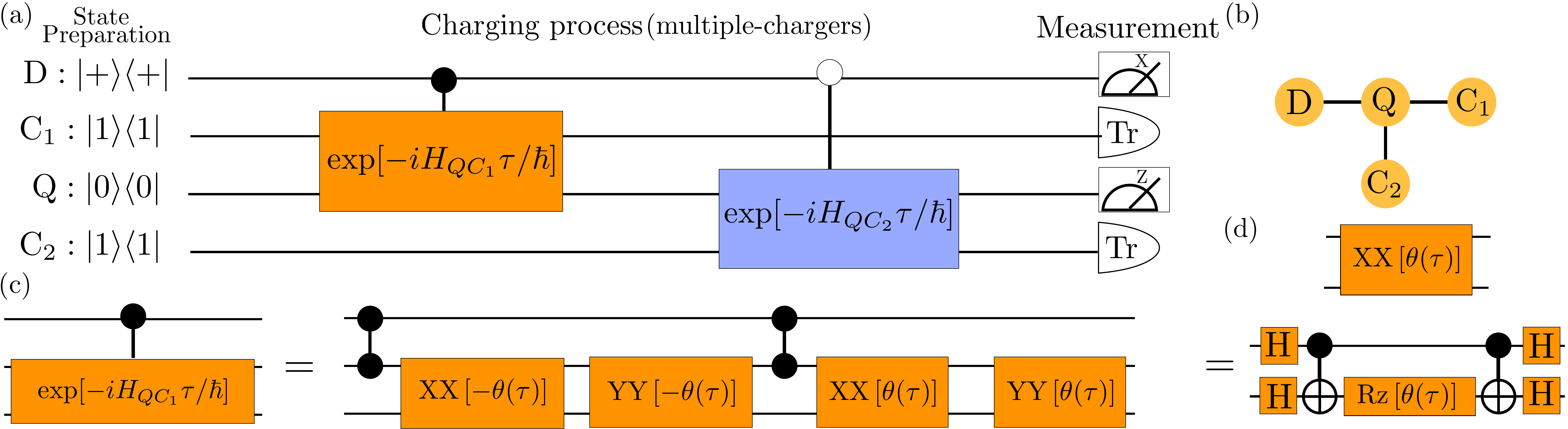}
\caption{(a) Quantum circuit for multiple-chargers protocol. Here, $D,Q,C_1,C_2$ represents the control qubit, battery qubit, first charger and second charger, respectively. (b) Decomposition of a controlled unitary in (a). (c) Qubit configuration used on ibmq$\_$algiers. (d) Decomposition of a $XX$ gate into CNOT gates.}
\label{independent circuit}
\end{figure*}

The quantum circuit for the multiple-charging setup is described by Fig.~\ref{independent circuit} (a). The circuit consists of 4 qubits, representing the control qubit $D$, the quantum battery $Q$, and the two charging cavities $C_1$ and $C_2$, respectively. The circuit can be divided into three parts: state preparation, charging process, and measurements on the control qubit and quantum battery. In the state preparation part, the qubits are prepared in the initial state specified in  Eq.~(\ref{initial_independent}) using single-qubit gates. The charging process involves the utilization of two controlled-unitary gates to simulate the simultaneous charging of the qubit by the two cavities through Jaynes-Cummings interactions. In Fig.~\ref{independent circuit} (b), we present the decomposition of the controlled--unitaries into bit-flip ($X$) gates, controlled-z gates ($CZ$), and Ising coupling gates [$XX(\theta)$ and $YY(\theta)$], defined as follows: 
\begin{equation}
\begin{aligned}
    &X = \hat{\sigma}_x,\\
    &CZ = \proj{0}\ts \id + \proj{1}\ts \hat{\sigma}_z, \\
    &XX\left(\theta\right)=\cos(\theta/2)\id\ts\id-i\sin(\theta/2)\hat{\sigma}_x\ts\hat{\sigma}_x, \\
    &YY\left(\theta\right)=\cos(\theta/2)\id\ts\id-i\sin(\theta/2)\hat{\sigma}_y\ts\hat{\sigma}_y.
\end{aligned}
\end{equation}

\begin{figure}[!htbp]
\includegraphics[width=1\columnwidth]{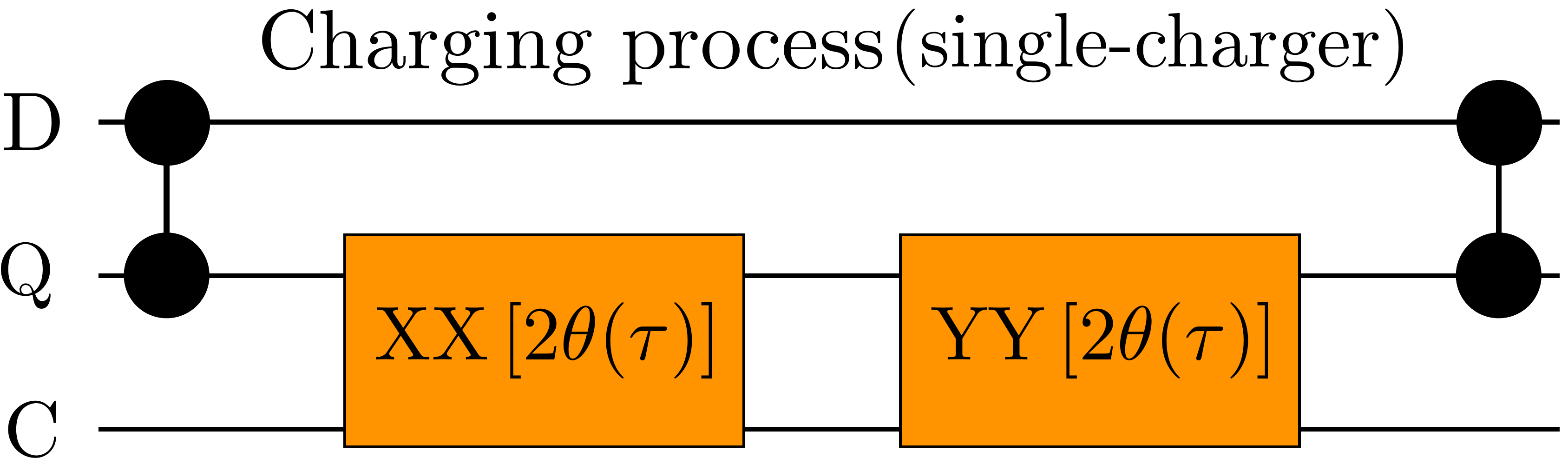}
\caption{Charging process of the single-charger setup, where the coupling strengths of the two superposed trajectories have the same magnitude but are out of phase.}
\label{same circuit}
\end{figure}

Here, we map the charging time $\tau$ into the angle $\theta$ using the following relation
\begin{equation}
    \theta(\tau) =\omega_c \lambda \tau / 2.
\end{equation}
Finally, in the measurement part, we measure the control qubit $D$ in the x-direction, aligned with the projectors described by Eq.~\eqref{eq:same projectors}. Furthermore, as indicated in Eq.~\eqref{eq:id post states}, the QB's post-measurement states are diagonalized under the energy eigenstates. Consequently, we can only measure $Q$ in the z-direction to determine the stored energy as well as the ergotropy. We utilize the ibmq$\_$algiers and IonQ-Aria 1 devices. Note that the qubit configuration for ibmq$\_$algiers is illustrated in Fig.~\ref{independent circuit} (c), while the qubits in IonQ-Aria 1 are fully connected. Also, since the Ising coupling gates are not native gates for the IBMQ device, we need to further decompose them into CNOT gates, which is shown in Fig.~\ref{independent circuit} (d). Consequently, the circuits for the IBMQ and IonQ devices consist of 20 and 12 two-qubit gates, respectively. Here, we use the Hadamard ($H$) gate and the rotation-z [$R_z(\theta)$] gate, which are defined by 
\begin{equation}
    \begin{aligned}
        &H = \frac{1}{2}
            \begin{pmatrix}
                1 & 1\\
                1 & -1
            \end{pmatrix},\\
        &R_z(\theta) = \exp(-i \hat{\sigma}_z \theta/2).
    \end{aligned}
\end{equation}
For the single-charger setup, the circuit model for the charging process is presented in Fig.~\ref{same circuit}, which consists of 6 and 4 two-qubit gates in the circuits for the IBMQ and the IonQ devices, respectively.

\begin{figure}[!htbp]
\includegraphics[width=1\columnwidth]{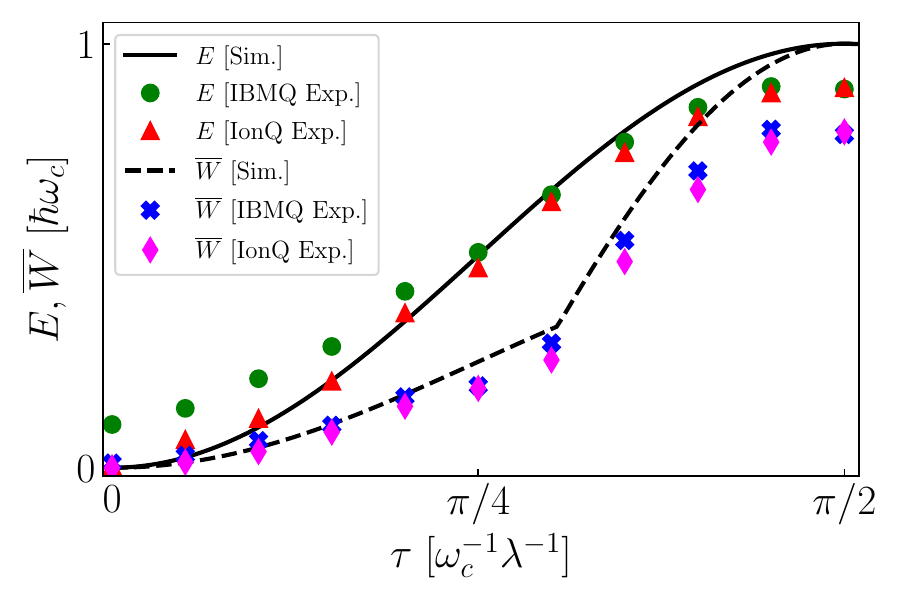}
\caption{The stored energy $E$ and average ergotropy $\overline{W}$ (both in units of $\hbar \omega_c$) on time $\tau$ (in units of $1/\omega_c$) for $\lambda = 0.05, N=2$. The black solid and dashed curves represent the stored energy and average ergotorpy predicted by numerical simulations. The green circles and blue “x"s  represent experimental results performed on ibmq$\_$algiers. The red triangles and magenta diamonds represent experimental results performed on IonQ Aria 1. Each data point is obtained after averaging 1000 experimental repetitions.}
\label{fig:independent experiment}
\end{figure}

\begin{figure}[!htbp]
\includegraphics[width=1\columnwidth]{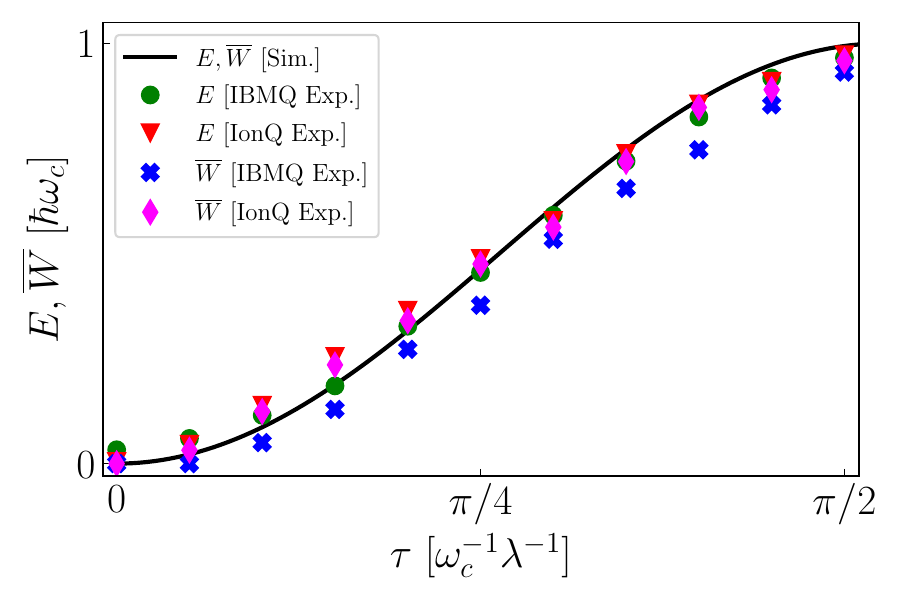}
\caption{The stored energy $E$ and average ergotropy $\overline{W}$ (both in units of $\hbar \omega_c$) on time $\tau$ (in units of $1/\omega_c$) with $r_1=0.1l$ and $r_2=0.9l$. The black curve represents the stored energy and average ergotropy predicted by numerical simulations. The green circle and blue ‘‘x"s represent the experimental results performed on ibmq$\_$algiers. The red triangles and magenta diamonds represent experimental results performed on IonQ Aria 1. Each data point is obtained after averaging 1000 experimental repetitions.}
\label{fig:same experiment}
\end{figure}

Figures~\ref{fig:independent experiment} and \ref{fig:same experiment} illustrate the results obtained from the two protocols, with each data point representing the average of 1000 experiment repetitions. The experimental results demonstrate a notable increase in average ergotropy, aligning with the theoretical predictions. Furthermore, we can observe that the deviation between the experimental and theoretical results is correlated by the circuit size, primarily determined by the number of two-qubit gates involved. Therefore, comparing the results from the IonQ and IBMQ devices, we find that the experimental data from the IonQ device exhibit a closer match to the theoretical curves compared to those from the IBMQ device. In addition, the errors associated with the single-charger protocol are smaller in magnitude than those of the multiple-chargers protocol.

\section{Summary and Outlook \label{summary}}
In this work, we utilize superposition of trajectories to propose two charging protocols for quantum batteries (QBs), leading to improved storage of extractable work. The first protocol, called the multiple-charger protocol, allows for simultaneous interaction between the QB and multiple chargers. Leveraging the interference effect, we observe an increase in extractable work, attributed to the Zeno-like state freezing effect. The second protocol, referred to as the single-charger protocol, enables the QB to interact with a single charger from multiple positions concurrently. We demonstrate that the collective interference effect generated by this setup allows the QB to achieve the maximum extractable work throughout the entire charging period. Moreover, we investigate the circuit implementations utilizing IonQ and IBMQ devices, providing experimental data that further support the enhanced extractable work, thus validating our theoretical predictions.

As a possible future direction, we could extend our charging protocols to a related framework called indefinite causal order~\cite{Rubino17,Goswami18,Ebler18,Zhao20,Loizeau20,Chiribella21}, which allows for the control of quantum operation ordering through a quantum switch. Based on this framework, we could consider the scenarios, where the ordering of the charging process becomes indefinite~\cite{Simonov22}. This exploration could shed new light on the potential benefits and implications of incorporating indefinite causal order into our proposed protocols, further advancing the field of quantum battery charging.


\section*{Acknowledgement}
We acknowledge the NTU-IBM Q Hub and the IBM quantum experience for providing us a platform to implement the experiment. This work is supported by the National Center for Theoretical Sciences and National Science and Technology Council, Taiwan, Grant Nos. MOST 111-2123-M-006-001and NSTC 111-2627-M-006-008.
\appendix

\section{Effects of the choice of projectors on the post-measurement  quantum battery states}\label{irrelevant projectors}
Here, we prove that the choice of projectors for $\ket{\xi_{k\neq1}}\bra{\xi_{k\neq1}}_{D}$ is irrelevant to the post-measurement states $\sigma_{k\neq 1}(\tau)$ as long as they are orthonormal to $\ket{\xi_{k=1}}\bra{\xi_{k=1}}_{D}$. 


Let us start from Eq.~(\ref{independent final eq}), which can be expanded into
\begin{equation}\label{ap:post-measurement battery state}
\begin{aligned}
    &\sigma_k(\tau)= \\
    &\frac{1}{N}\sum^N_{j,f=1}\braket{\xi_k|j}\braket{f|\xi_k}\text{Tr}_C\left[\ket{\phi_j(\tau)}\bra{\phi_f(\tau)}_{QC}\right]
\end{aligned}
\end{equation}

We switch to the interaction picture, where the interaction Hamiltonian reads
\begin{equation}
    H'_{QC_j}(\tau)= e^{\frac{i}{\hbar}(H_Q+H_{C_j})\tau}H'_{QC_j}e^{-\frac{i}{\hbar}(H_Q+H_{C_j})\tau}.
\end{equation}
%
The time-dependent part of post-measurement states can be generally expressed by
\begin{equation}\label{ap: interaction begin}
\begin{aligned}
    &\ket{\phi_{j}\left(\tau\right)}_{QC}=\sum_{n=0}^\infty\alpha_{n}(\tau)\ket{n_j},
\end{aligned}
\end{equation}
with
\begin{equation}
    \ket{n_j}=(\hat{\sigma}_x)^{n}\ket{e}\ts\frac{(\hat{a}_j^{\dag})^{n}\hat{a}_j}{\sqrt{n!}}\bigotimes_{m=1}^N\ket{1}_{C_m}.
\end{equation}
Note that in the following analysis, we show the explicit expression of the time-dependent coefficient $\alpha_{n}(\tau)$ does not affect the result. Thus, we will keep them unspecified. Now, we can show that
\begin{equation}\label{ap: interaction end}
\begin{aligned}
    &\text{Tr}_C\left[\ket{\phi_{j}\left(\tau\right)}\bra{\phi_{f}\left(\tau\right)}_{QC}\right] \\
    &=\begin{cases}
        \sum_{n=0}^\infty{|\alpha_{n}(\tau)|}^2\sigma_x^{n}\ket{e}\bra{e}\sigma_x^{n}& \text{for~~}j=f \\
        {|\alpha_1(\tau)|}^2\ket{g}\bra{g},&\text{for~~}j\neq f.
    \end{cases}
\end{aligned}
\end{equation}
Inserting this into Eq.~(\ref{ap:post-measurement battery state}), we can obtain
\begin{equation}
\begin{aligned}
    &\sigma_k(\tau)= \sum_{n=0}^\infty{|\alpha_n(\tau)|}^2\sigma_x^n\ket{e}\bra{e}\sigma_x^n\frac{1}{N}\sum_{j=f}^N{|\braket{\xi_k|j}|}^2 \\
    &~~~~~~~~~+{|\alpha_1(\tau)}|^2\ket{g}\bra{g}\frac{1}{N}\sum_{j\neq f}^N\sum_{f=1}^N\braket{\xi_k|j}\braket{f|\xi_k}.
\end{aligned}
\end{equation}
For the case of $k=1$, according to Eq.~\eqref{eq:projector}, we can obtain
\begin{equation}
\begin{aligned}
    \braket{\xi_{k=1}|j}=\frac{1}{\sqrt{N}}~~~~\forall~j=1,\cdots, N.
\end{aligned}
\end{equation}
Therefore, the corresponding post-measurement state can be written as 
\begin{equation}
\begin{aligned}
    \sigma_{k=1}(\tau)=&\frac{1}{N}\sum_{n=0}^\infty{|\alpha_n(\tau)|}^2\sigma_x^{n}\ket{e}\bra{e}\sigma_x^{n}\\ 
    &+\frac{N-1}{N}{|\alpha_1(\tau)|}^2\ket{g}\bra{g}.
\end{aligned}
\end{equation}
Note that in the asymptotic limit ($N\rightarrow\infty$), we observe $\sigma_{k=1}(\tau) \propto \proj{g}$, indicating the Zeno-like state freezing effect~\cite{Lin2022}.

Recall that for the case of $k\neq 1$, we require the projectors to be orthonormal to that of $k=1$, i.e., $\braket{\xi_{k\neq 1}|\xi_{k=1}}=0$, implying that the ket $\ket{\xi_{k\neq 1}}$ can be expressed by
\begin{equation}
\begin{aligned}
    &\ket{\xi_{k\neq 1}}=\sum_{j=1}^N \beta_{k,j}\ket{j}_D, \\&
    \text{with~}\sum_{j=1}^N \beta_{k,j}=0~~~\forall~k\neq 1.
\end{aligned}
\end{equation}
Therefore, the post-measurement states can be written as\begin{equation}
\begin{aligned}
    \sigma_{k\neq 1}(\tau)=&\frac{1}{N}\sum_{n=0}^\infty{|\alpha_n(\tau)|}^2\sigma_x^{n}\ket{e}\bra{e}\sigma_x^{n} \\
    &-\frac{1}{N}{|\alpha_1(\tau)|}^2\ket{g}\bra{g}.
\end{aligned}
\end{equation}
This concludes the proof that the post-measurement states are independent of the explicit expression of the coefficients $\beta_{k,j}$ associated with the projectors.

\end{document}